\documentclass{emulateapj}
\usepackage{apjfonts}

\slugcomment{Submitted to ApJ Letters on 10/26/2006; accepted on 11/14/2006}

\shorttitle{Mid-IR galaxy classification}
\shortauthors{Spoon et al.}

\begin{document}

\title{Mid-IR galaxy classification based on silicate 
obscuration and PAH equivalent width}

\author{H.W.W. Spoon\altaffilmark{1,2}}
\email{spoon@astro.cornell.edu}
\author{J.A. Marshall\altaffilmark{1}}
\author{J.R. Houck\altaffilmark{1}}
\author{M. Elitzur\altaffilmark{3}}
\author{L. Hao\altaffilmark{1}}
\author{L. Armus\altaffilmark{4}}
\author{B.R. Brandl\altaffilmark{5}}
\author{V. Charmandaris\altaffilmark{6,7}}

\altaffiltext{1}{Cornell University, Astronomy Department, Ithaca, NY 14853}
\altaffiltext{2}{Spitzer Fellow}
\altaffiltext{3}{Department of Physics and Astronomy, University of
             Kentucky, Lexington, KY 40506}
\altaffiltext{4}{Caltech, Spitzer Science Center, MS 220-6, Pasadena, CA 91125}
\altaffiltext{4}{Leiden Observatory, PO Box 9513, 2300 RA Leiden,
             The Netherlands}
\altaffiltext{6}{Department of Physics, University of Crete,   
             GR-71003, Heraklion, Greece}
\altaffiltext{7}{IESL / Foundation for Research and Technology-Hellas, 
             PO Box 1527, GR-71110, Heraklion, Greece, and Chercheur 
             Associ\'e, Observatoire de Paris, F-75014,  Paris, France}

\begin{abstract}
We present a new diagnostic diagram for mid-infrared spectra of 
infrared galaxies based on the equivalent width of the 6.2\,$\mu$m PAH
emission feature and the strength of the 9.7\,$\mu$m silicate
feature. Based on the position in this diagram we classify galaxies 
into 9 classes ranging from continuum-dominated AGN hot dust spectra
and PAH-dominated starburst spectra to absorption-dominated spectra
of deeply obscured galactic nuclei.
We find that galaxies are systematically distributed along 
two distinct branches: one of AGN and starburst-dominated spectra 
and one of deeply obscured nuclei and starburst-dominated spectra. 
The separation into two branches likely reflects a fundamental
difference in the dust geometry in the two
sets of sources: clumpy versus non-clumpy obscuration.
Spectra of ULIRGs are found along the full length of both branches, 
reflecting the diverse nature of the ULIRG family.
\end{abstract}


\section{Introduction}

Over the last decade several diagnostic diagrams have been 
proposed to quantify the contribution of star formation and
AGN activity to the infrared luminosity of infrared galaxies
based on mid-infrared (to far-infrared) continuum slope,
PAH line-to-continuum ratio, PAH to far-infrared luminosity
ratio and the ratio of a high to a low ionization forbidden
line such as [Ne{\sc v}]/[Ne{\sc ii}] 
\citep{genzel98,lutz98,laurent00,peeters04,dale06,sturm06}.
However, none of these diagrams takes into account the effects 
of strong obscuration of the nuclear power source.

With the advent of the Infrared Spectrograph \citep[{\it IRS};][]{houck04} 
on board the {\it Spitzer} Space Telescope \citep{werner04} 
astronomers have been handed a powerful tool 
to study the 5--37\,$\mu$m range for a wide range of galaxy types
at an unprecedented sensitivity. This enables for the first time
a systematic study of a large number of galaxies over the wavelength 
range in which amorphous silicate grains have strong opacity peaks 
due to the Si--O stretching and the O--Si--O bending modes centered 
at 9.7 and 18\,$\mu$m, respectively.

Here we will introduce the strength of the 9.7\,$\mu$m silicate feature
as a tool to distinguish between different dust geometries in the 
central regions of (ultra-)(luminous) infrared galaxies as part
of a new diagnostic diagram and mid-infrared galaxy classification 
scheme.

\section{Observations and data reduction}

The results presented in this paper are based on both {\it Spitzer IRS}
and {\it ISO-SWS} observations. The core sample is formed by the IRS 
GTO ULIRG sample (PID 105; J.R.\,Houck PI), which comprises $\sim$100 
ULIRGs in the redshift range 0.02 $< z <$ 0.93.
This sample is compared to samples of AGN and starburst templates
from the IRS GTO programs 14 \citep{brandl06,weedman05} and 96
and to selected AGNs from the GTO programs 82 and 86 \citep{shi06}.
Additional IRGs, LIRGs and ULIRGs were taken from program 159 
\citep[NGC\,1377;][]{roussel06} and from the Spitzer DDT program
1096. Archival {\it ISO-SWS} spectra of starburst nuclei were taken from 
\cite{sturm00}.

The {\it Spitzer} observations were made with the Short-Low (SL) and 
Long-Low (LL) modules of the {\em IRS}. The spectra were extracted 
from DROOP level images provided by the Spitzer Science Center 
(34 using pipeline version S11.0.2, 131 using pipeline version
S14.0) and background-subtracted 
by differencing the first and second order apertures. The spectra
were calibrated using the {\em IRS} standard stars HD\,173511 
(5.2--19.5\,$\mu$m) and $\xi$\,Dra (19.5--38.5\,$\mu$m).
After flux calibration the orders were stitched to LL order 1, requiring 
order-to-order scaling adjustments of typically less than 10\%.

\begin{figure*}
\epsscale{1.15}
\plotone{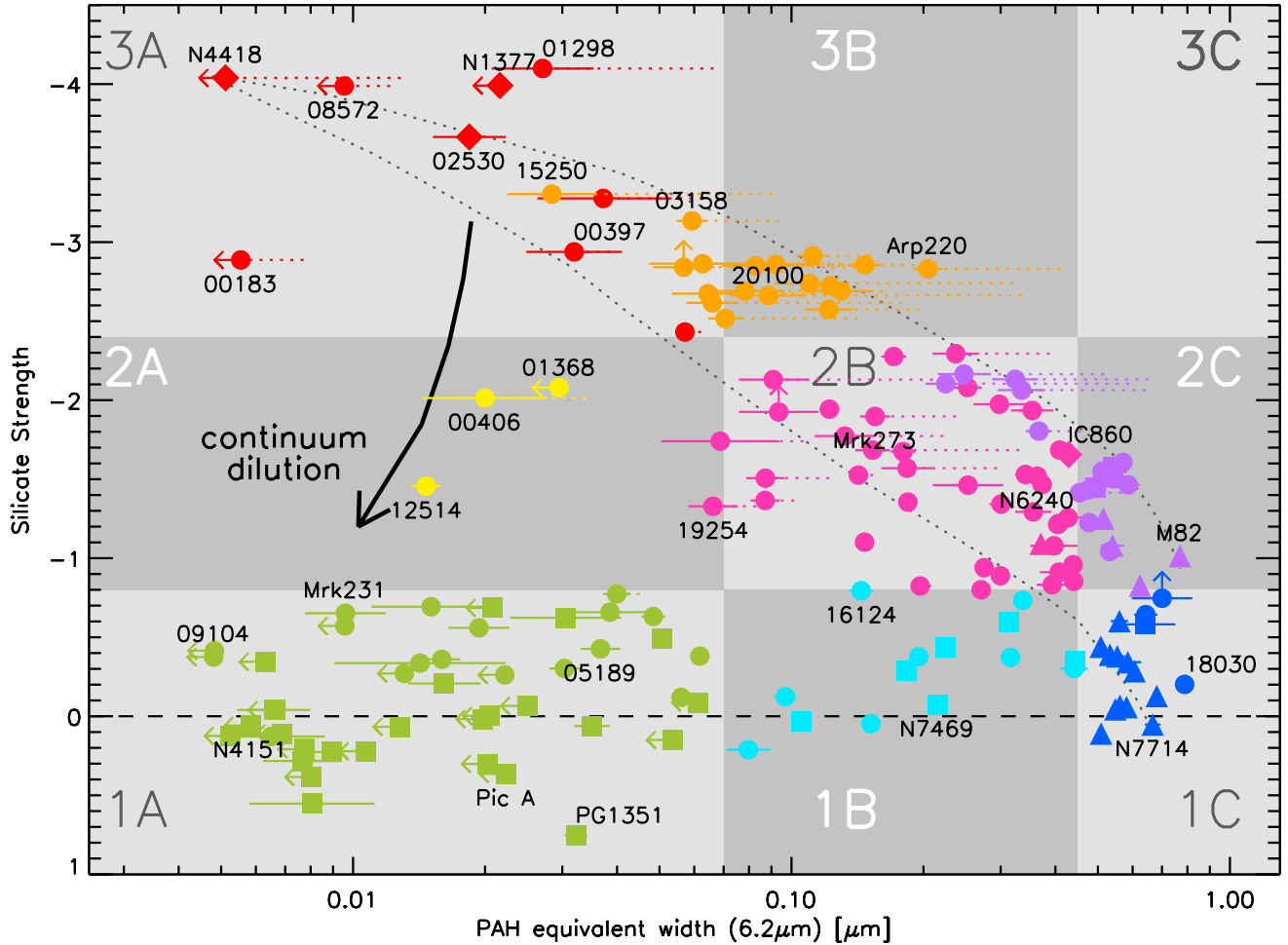}
\caption{Diagnostic plot of the equivalent width of the 6.2\,$\mu$m 
PAH emission feature versus the 9.7\,$\mu$m silicate strength.
Upper and lower limits are denoted by arrows. The galaxy spectra 
are classified into 9 classes (identified by 9 shaded rectangles) 
based on their position in this plot. 
Colors are used to distinguish the various classes. 
From class 1A to 1B, 1C, 2A, 2B, 2C, 3A and 3B the colors used are: 
{\it green, cyan, dark blue, yellow, pink, purple, red} and {\it orange}. 
After class assignment the PAH equivalent widths were corrected for
the effect of 6\,$\mu$m water ice absorption on the 6.2\,$\mu$m
continuum. The extent of the individual corrections are indicated by 
dotted horizontal lines.
The two dotted black lines are mixing lines between the spectrum of 
the deeply obscured nucleus of NGC\,4418 and the starburst
nuclei of M\,82 and NGC\,7714, respectively.
Galaxy types are distinguished by their plotting symbol:
{\it filled circles}: ULIRGs and HyLIRGs; {\it filled triangles}: 
starburst galaxies; {\it filled squares}: Seyferts and QSOs; 
{\it filled diamonds}: other infrared galaxies.
\label{fig1}}
\end{figure*}

\section{Analysis}

For all the spectra in our sample we have measured the equivalent 
width of the 6.2\,$\mu$m PAH emission feature as well as the 
strength of the 9.7\,$\mu$m silicate feature and plotted the two 
quantities in a diagnostic diagram (Fig.\,\ref{fig1}).

The flux in the 6.2\,$\mu$m PAH emission band is measured by
integrating the flux above a spline interpolated local continuum 
from 5.95--6.55\,$\mu$m. The equivalent
width (EW) of the PAH feature is then obtained by dividing the 
integrated PAH flux by the interpolated continuum flux density 
below the peak ($\sim$ 6.22\,$\mu$m) of the PAH feature.

The apparent strength of the 9.7\,$\mu$m silicate feature
is inferred by adopting a local mid-infrared continuum 
and evaluating the ratio of observed flux density (f$_{\rm obs}$) to 
continuum flux density (f$_{\rm cont}$) at 9.7\,$\mu$m and defining 
\begin{equation}
S_{sil} = \ln \frac{f_{\rm obs} (9.7\,\mu m)}{f_{\rm cont} (9.7\,\mu m)}
\end{equation}
For sources with a silicate absorption feature $S_{sil}$ can be 
interpreted as the negative of the apparent silicate optical depth.

\begin{figure}
\epsscale{1.15}
\plotone{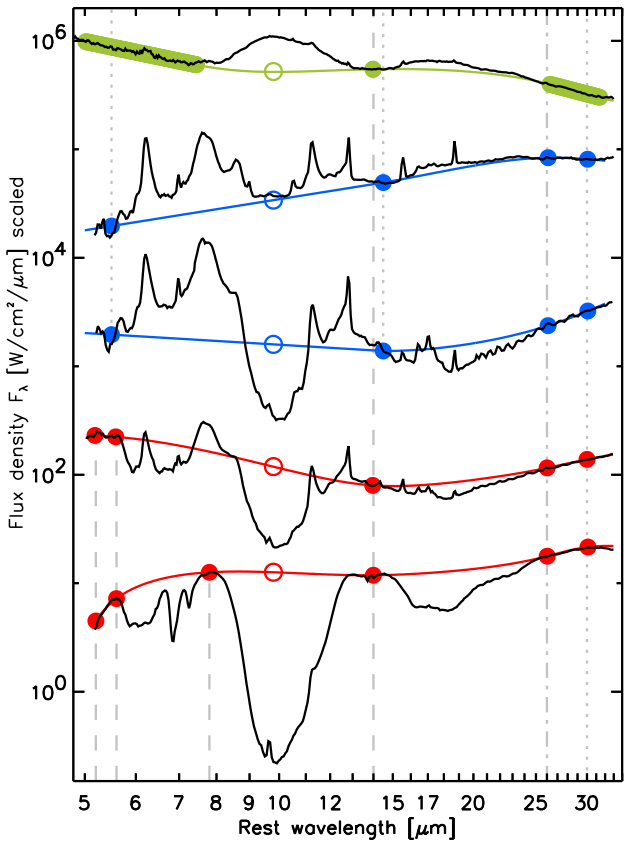}
\caption{Comparison of several methods used to determine the 
apparent 9.7\,$\mu$m silicate optical depth. The three methods
differ by the way the local mid-infrared continuum is defined.
For continuum-dominated spectra (top) simple logarithmic 
linear fits are made to the 5.0--7.5\,$\mu$m and 26.1--31.5\,$\mu$m
Jy-spectrum. These are spline interpolated over the 7.5--26\,$\mu$m
range, using a continuum point at 14.0\,$\mu$m to guide the
spline in the inter-silicate-feature range. Pivots and fit
are shown in {\it green}. 
For PAH-dominated spectra (2nd and 3rd spectrum)
the 5.5--14.5\,$\mu$m continuum is estimated from a power law
interpolation over this range. From 14.5 to 26.0\,$\mu$m this
power law is continued as a spline function with a curvature 
determined by continuum pivots at 26.0 and 30.0\,$\mu$m. Pivots 
and fit are shown in {\it blue}. 
For absorption-dominated spectra (4th and 5th spectrum)
the local 5.6--26\,$\mu$m continuum is determined from a spline
interpolation to continuum pivots at 5.2, 5.6, 14.0 and 
26.0\,$\mu$m. An additional continuum pivot at 7.8\,$\mu$m
is included for spectra with very little PAH emission (e.g.
bottom spectrum). Pivots and fit are shown in {\it red}. 
In all three cases the 9.7\,$\mu$m silicate strength 
is computed by evaluating the ratio of observed and interpolated 
continuum at its peak. From top to bottom, the five spectra 
shown are the IRS low-resolution spectra of PG\,1351+640, NGC\,7714,
NGC\,3628, UGC\,5101 and NGC\,4418.
\label{fig2}}
\end{figure}

Given the great diversity among our mid-infrared galaxy spectra, 
there is no ``one-size-fits-all'' procedure to define the local 
continuum in all spectra. We therefore developed separate methods for 
three distinctly different types of mid-infrared galaxy spectra: 
continuum-dominated spectra, PAH-dominated spectra and 
absorption-dominated spectra. These methods are illustrated in 
Fig.\,\ref{fig2} and described in its caption.
We are forcing the local continuum to touch down at
14.0--14.5\,$\mu$m, because the ISM dust cross-section 
decreases between the two silicate peaks in this region. 
Detailed radiative transfer calculations verify that 
our interpolation procedure properly reproduces the 
emission that would be generated by dust stripped of 
its silicate features (Sirocky et al. in preparation).

\subsection{Galaxy classification}

\begin{figure}
\epsscale{1.15}
\plotone{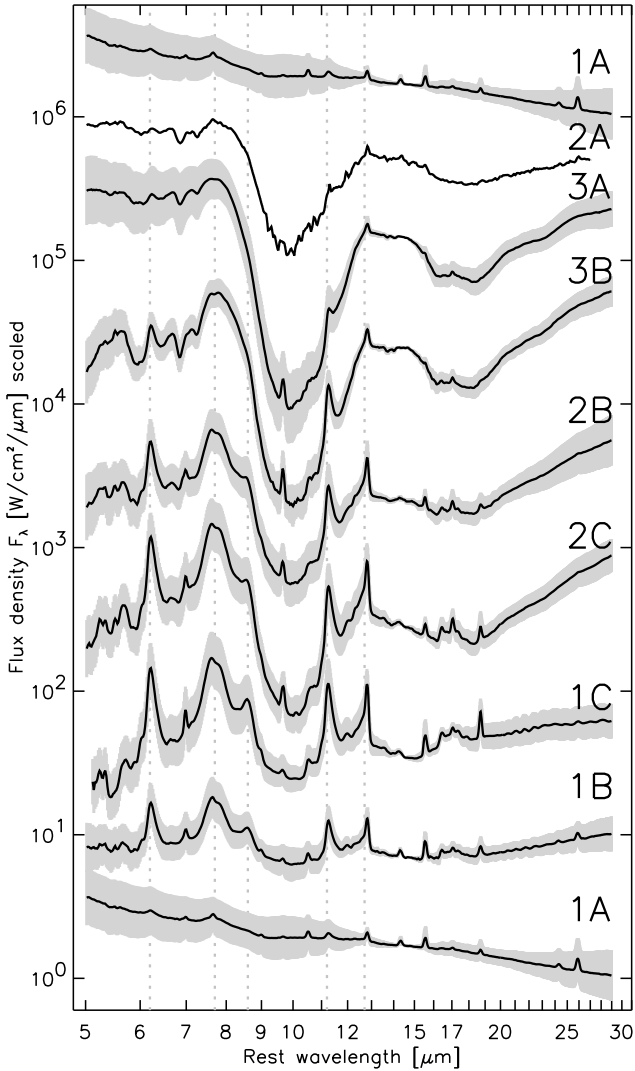}
\caption{Average spectra for eight of the classes of galaxy
spectra defined in Fig.\,\ref{fig2}, along with the 1-$\sigma$ 
dispersion range ({\it grey} shading). Most of the dispersion
is due to the overall slope of the spectrum. Spectra at the 
bottom of the
figure (classes 1A and 1B) are dominated by hot dust emission, 
spectra in the middle (classes 1C and 2C) by PAH emission and
spectra towards the top by absorption features 
(classes 3B, 3A and 2A).
Vertical dotted lines denote the central wavelengths 
of the family of PAH emission features at 6.2, 7.7, 8.6, 11.2 and 
12.7\,$\mu$m. 
\label{fig3}}
\end{figure}

The galaxy spectra in Fig.\,\ref{fig1} are classified into
9 different classes based on their 6.2\,$\mu$m PAH EW and
9.7\,$\mu$m silicate strength. The parameter space
covered by the various classes is indicated by shaded 
rectangles in Fig.\,\ref{fig1}. Average mid-infrared 
spectra for the eight populated classes are shown in 
Fig.\,\ref{fig3}.
The average spectra were constructed by normalizing all spectra 
to unity at 14.5\,$\mu$m flux before the averaging process. 
In order to maximize the signal-to-noise (S/N) of the average 
spectra, low S/N spectra were discarded from the process.
Below we describe the 8 average spectra in the order (bottom 
to top) they are presented in Fig.\,\ref{fig3}. The ninth
class, 3C, is not populated.

The class 1A spectrum is characterized by a nearly featureless
hot dust continuum with a very weak silicate absorption feature 
at 9.7\,$\mu$m. The class 1B spectrum differs from the 
class 1A spectrum by clearly showing the family of PAH emission 
features at 6.2, 7.7, 8.6, 11.2, 12.7 and 17.3\,$\mu$m on top 
of a hot dust continuum. This hot dust continuum is nearly 
absent in the class 1C spectrum, allowing the PAH emission 
features to dominate the mid-infrared spectral appearance. 
Silicate absorption at
9.7\,$\mu$m becomes noticable in the class 2C spectrum as 
an increased depth of the depression between the 6--9 and 
11--13\,$\mu$m PAH emission complexes. Another marked 
difference with the class 1C spectrum is the steepening
of the 20--30\,$\mu$m continuum and the appearance of an
18\,$\mu$m silicate absorption feature. In the class 2B 
spectrum the PAH features appear weaker than in the 
class 2C spectrum. In addition, the 
spectrum starts to show a 6\,$\mu$m water ice and a 
6.85\,$\mu$m aliphatic hydrocarbon absorption band.
In the class 3B spectrum these absorption features reach
their maximum depths, while the 9.7 and 18\,$\mu$m silicate 
features continue to increase in depth up to class 3A.
Absorption features of {\it crystalline} silicates \citep{spoon06}
appear in the spectra of the classes 3B and 3A at 16, 19 and 23\,$\mu$m. 
Equivalent widths of PAH emission features and emission 
lines decrease from class 2B to 3B and 3A. Note especially
the change in shape of the 7.7\,$\mu$m PAH feature as
it first broadens and then disappears going from class 
2B to 3B and 3A. Finally, the class 2A spectrum differs 
from the class 3A spectrum mainly by a clearly lower 
apparent depth of the 9.7 and 18\,$\mu$m silicate features.
Note that the spectral structure in individual spectra may 
differ substantially from the average properties of the
classes. In Fig.\,\ref{fig3} this is represented by the
1-$\sigma$ dispersion ranges around the individual average 
spectra.

\subsection{Galaxy distribution}

The galaxies shown in Fig.\,\ref{fig1} are color-coded according 
to their galaxy classification. However, their positions may differ 
from their original assignments, as in Fig.\,\ref{fig1} the 6.2\,$\mu$m 
PAH EW has been corrected for the effect of 6\,$\mu$m water ice 
absorption on the 6.2\,$\mu$m continuum through substitution of 
the observed 6.2\,$\mu$m continuum by the 6.2\,$\mu$m continuum 
that was defined to infer the 9.7\,$\mu$m silicate strength. 
The ice-correction is justified assuming that the ice is part of 
the obscuring medium {\it behind} the PAH emitting region.
The correction is applied only to spectra clearly showing the 
imprint of water ice absorption, as found among the spectra of 
classes 2B, 2C, 3A and 3B. 
The extent of the correction for individual spectra is indicated 
in Fig.\,\ref{fig1} by horizontal dotted lines. 

The ULIRGs, Seyferts, quasars and starburst galaxies shown in
Fig.\,\ref{fig1} are not distributed randomly through the diagram.
Instead, the galaxies appear to be distributed along two branches.
One extending horizontally from the positions of the prototypical
Seyfert-1 nucleus NGC\,4151 \citep{weedman05} to the prototypical 
starburst nucleus NGC\,7714 \citep{brandl04}, the other branch 
extending diagonally from the 
prototypical deeply obscured nucleus of NGC\,4418 \citep{spoon01} 
to the starburst nuclei of M\,82 \citep{sturm00} and NGC7714.
Very few sources are found above and to the right of the diagonal 
branch and in between the two branches (the class 2A domain).

The three extremes distinguish between AGN-heated hot dust 
spectra (class 1A), PAH-dominated spectra (classes 1C and 2C)
and absorption-dominated spectra (class 3A).  The galaxies lined 
up along the two branches in between the extremes show signatures
of both extremes to varying proportions. The galaxies along the
horizontal branch may be thought of as combinations of AGN and 
starburst activity, while the galaxies along the diagonal branch 
may be thought of as intermediate stages in between 
a fully obscured galactic nucleus and an unobscured nuclear 
starburst. The latter is illustrated by the two dotted lines 
in Fig.\,\ref{fig1}, which represent mixing lines between 
the spectrum of NGC\,4418 on one extreme and either NGC\,7714 or
M\,82 on the other. The large majority of galaxies on the diagonal 
branch fall in between these mixing lines. One notable exception 
is Arp\,220.

Relatively few galaxies are found in the section between the two 
branches: the domain of the class 2A sources. Spectroscopically, 
there are several simple scenarios which will populate this class 
from adjacent classes.
A class 3A galaxy can be turned into a class 2A galaxy by 
adding a featureless hot dust continuum to the class 3A absorbed 
continuum spectrum. The main effect of this continuum dilution is 
the filling in of the deep 9.7\,$\mu$m silicate feature, resulting 
in a less pronounced feature. Increasing the dilution
not only further decreases the depth of the silicate feature
but also increases the continuum at 6.2\,$\mu$m, resulting in 
both a less negative silicate strength and a lower 6.2\,$\mu$m 
PAH EW. The effect of this type of continuum dilution is illustrated 
in Fig.\,\ref{fig1} by a curved arrow.
Alternatively, a class 1A galaxy may be turned into class 2A 
galaxy by passing its mid-infrared spectrum through a foreground 
cold dust screen. The path that a galaxy will take is straight up, 
since the PAH EW is not affected by cold foreground extinction.

The presence of foreground dust or dust mixed in within the PAH
emitting region may also explain the difference between class 1C 
and 2C starburst galaxies, and the difference between the 
classification of the prototypical starburst galaxies NGC\,7714 
and M\,82 in particular. In the latter case this may be primarily 
an orientation effect, as the nuclear starburst in NGC\,7714 is 
observed face-on, whereas the starburst ring in M\,82 is seen under 
an inclination of $\sim$80 degrees.

Finally, the complete absence of galaxy spectra classified as 3C must imply 
that strong, pure foreground, cold dust obscuration is not common 
in starburst nuclei. The obscuring dust instead contributes to the
continuum underlying the 6.2\,$\mu$m PAH feature, reducing its EW.

\section{Discussion and conclusions}

We have constructed a diagnostic diagram of two mid-infrared
spectral quantities, the equivalent width of the 6.2\,$\mu$m 
PAH feature and the apparent 9.7\,$\mu$m silicate strength, for 
the purpose of classifying infrared galaxies according to their 
mid-infrared spectral shape (Fig.\,\ref{fig1}).

The large majority of galaxies ($>$90\%) are dispersed around 
two branches: a horizontal one extending from 
continuum-dominated spectra to PAH-dominated spectra and a 
diagonal one spanned between absorption-dominated spectra 
and PAH-dominated spectra.
Seyfert galaxies and quasars (square plotting symbols in 
Fig.\,\ref{fig1}) are found exclusively on the horizontal branch
characterized by the absence of pronounced silicate features.
Starburst galaxy spectra (triangular plotting symbols) are 
concentrated toward the extreme right 
tip of the two branches, the locus of PAH-dominated spectra.
Of the ULIRGs and HyLIRGs in our sample (round plotting symbols) 
about a quarter occupy the same
space as the AGNs and starbursts in our sample. The large
majority, however, are distributed along the diagonal branch,
characterized by increasingly apparent silicate absorption
and less pronounced PAH emission features, with deeply
obscured galactic nuclei such as NGC\,4418 \citep{spoon01} 
and NGC\,1377 \citep{roussel06} as the only non-ULIRG, 
non-Seyfert end members on this branch.
Note that low-metallicity galaxies (not included in this
study) would be distributed along the horizontal branch.

The existence of two distinct regimes of apparent silicate 
absorption, as suggested by the two distinct branches in 
Fig.\,\ref{fig1}, is intriguing and likely points to clear 
differences in the nuclear dust distribution between galaxies 
on the two branches. Clumpy distributions produce only 
shallow absorption features \citep{nenkova02}, and are 
the likely explanation for the horizontal branch. The deep 
silicate absorption found on the slanted branch requires the 
nuclear source to be deeply embedded in a smooth distribution 
of dust that is both geometrically and optically thick 
\citep{levenson06}. The transition from clumpy geometry to 
one dominated by a smooth dust distribution is a possible 
explanation for the sources located between the branches.

The distribution of ULIRG spectra along the full length
of both the horizontal and the diagonal branch once again
illustrates the diverse nature of the ULIRG family. 
It, however, also raises the issue of ULIRG evolution.
For instance, what evolutionary path has taken 
IRAS\,08572+3915 to its extreme class 3A position in 
which the central power source is deeply buried? 
Did the interaction start with both nuclei classified
as class 1C/2C starburst galaxies, which then gradually 
moved up the diagonal branch as the interaction 
stengthened and more dust accumulated on the remnant
nuclei? And how will IRAS\,08572+3915 evolve from
there, once the obscuring screen breaks up and the hidden 
power sources are revealed? In case of a hidden starburst
the source likely will move back diagonally toward the 
starburst locus. However, if the dominant power source 
is an AGN, will the source cross into the sparsely 
populated class 2A regime, evolving directly toward the 
AGN-dominated class 1A regime, or will it first 
undergo a starburst before settling somewhere along 
the horizontal branch?
The sparse population of the class 2A domain either 
indicates that the crossing time is brief -- with
few ULIRGs caught in transformation -- or that deeply
obscured ULIRG nuclei mostly evolve to starbursts 
(first).

In a more extensive paper (Spoon et al. in preparation) 
we will present correlations of additional parameters 
within the diagram (e.g. mid-IR line ratios, L$_{\rm IR}$, 
L$_{\rm PAH}$/L$_{\rm IR}$,absorption features, nuclear 
separation, optical classification).


\acknowledgements

The authors wish to thank Jer\'onimo Bernard-Salas, 
Nancy Levenson, H\'el\`ene Roussel, Matthew Sirocky,
Alexander Tielens and Dan Weedman for discussions, and
Vandana Desai for help with the data reduction. 
Support for this work was provided by 
NASA through Contract Number 1257184 issued by the Jet Propulsion 
Laboratory, California Institute of Technology under NASA contract 
1407.  HWWS was supported under this contract through the 
Spitzer Space Telescope Fellowship Program.



\begin{thebibliography}{}

\bibitem[Brandl et al.(2004)Brandl et al.]{brandl04} 
         Brandl, B.R., et al., 2004, \apjs 154, 188

\bibitem[Brandl et al.(2006)Brandl et al.]{brandl06} 
         Brandl, B.R., et al., 2006, \apj, 653 (astro\_ph/0609024)

\bibitem[Dale et al.(2006)Dale et al.]{dale06} 
         Dale, D.A., et al., 2006, \apj, 646, 161

\bibitem[Genzel et al.(1998)Genzel et al.]{genzel98}
         Genzel, R., et al., 1998, \apj, 498, 579

\bibitem[Houck et al.(2004)Houck et al.]{houck04} 
         Houck, J.R., et al. 2004, \apjs, 154, 18

\bibitem[Laurent et al.(2000)Laurent et al.]{laurent00} 
         Laurent, O., et al., 2000, \aap, 359, 887

\bibitem[Levenson et al.(2006)Levenson et al.]{levenson06}
         Levenson, N.A., et al., 2006, \apjl, accepted

\bibitem[Lutz et al.(1998)Lutz et al.]{lutz98}
         Lutz, D., et al., 1998, \apjl, 505, L103

\bibitem[Nenkova, Ivezic \& Elitzur (2002)Nenkova, Ivezic \& Elitzur]{nenkova02}
Nenkova, M., Ivezic, Z., \& Elitzur, M. 2002, \apjl, 570, L9

\bibitem[Peeters et al.(2004)Peeters et al.]{peeters04}
         Peeters, E., Spoon, H.W.W., Tielens, A.G.G.M.,
         2004, \aap, 613, 986

\bibitem[Roussel et al.(2006)Roussel et al.]{roussel06} 
         Roussel, H., et al., 2006, \apj, 646, 841

\bibitem[Shi et al.(2006)Shi et al.]{shi06} 
         Shi, Y., et al., 2006, 
         \apj, in press (astro\_ph/0608645)

\bibitem[Spoon et al.(2001)Spoon et al.]{spoon01} 
         Spoon, H.W.W., et al., 2001, \aap, 365, L353

\bibitem[Spoon et al.(2006)Spoon et al.]{spoon06} 
         Spoon, H.W.W., et al., 2006, \apj, 638, 759

\bibitem[Sturm et al.(2000)Sturm et al.]{sturm00}
         Sturm, E., et al., 2000, \aap, 358, 481

\bibitem[Sturm et al.(2006)Sturm et al.]{sturm06}
         Sturm, E., et al., 2006, \apjl, in press (astro\_ph/0610772)

\bibitem[Weedman et al.(2005)Weedman et al.]{weedman05} 
         Weedman, et al., 2005, \apj

\bibitem[Werner et al.(2004)Werner et al.]{werner04} 
         Werner, M.W., et al., 2004, \apjs, 154, 1

\end{thebibliography}
\end{document}